\documentstyle[12pt]{article}
\def\beq{\begin{equation}}
\def\eeq{\end{equation}}
\def\bea{\begin{eqnarray}}
\def\eea{\end{eqnarray}}
\def\bq{\begin{quote}}
\def\eq{\end{quote}}

\def\be{\begin{equation}}
\def\ee{\end{equation}}
\def\ba{\begin{eqnarray}}
\def\ea{\end{eqnarray}}

\def\gappeq{\mathrel{\rlap {\raise.5ex\hbox{$>$}}
{\lower.5ex\hbox{$\sim$}}}}

\def\lappeq{\mathrel{\rlap{\raise.5ex\hbox{$<$}}
{\lower.5ex\hbox{$\sim$}}}}

\begin{document}
\pagestyle{empty}
\begin{flushright}
{CERN-TH/97-104}\\
\end{flushright}
\vspace*{5mm}
\begin{center}
{\bf A SCENARIO FOR CONTACT INTERACTIONS AT HERA}\\
\vspace*{1cm} 
{\bf Francesco CARAVAGLIOS} \\
\vspace{0.3cm}
Theoretical Physics Division, CERN \\
CH - 1211 Geneva 23 \\
\vspace*{2cm}  
{\bf ABSTRACT} \\ \end{center}
\vspace*{5mm}
\noindent
 The four fermion 
contact interactions, required to explain the anomalous HERA result,
could come from the exchange of new heavy (probably composite) resonances.
Depending on their charges and  quantum numbers, one gets different 
scenarios and  finds that many of these configurations are unsuitable.    
For example, new neutral resonances   seems to be disfavored 
by the data coming from the TEVATRON, LEP 2 and atomic parity violation.
 These experiments     allow only few helicity
 combinations  that cannot arise from neutral currents in a natural way.
On the contrary,   a global large symmetry $SU(8)\times SU(8)$ (which is
contained in $SU(16)$) embeds  some lepto-quarks of spin 1 that could give
suitable four fermion interactions (compatible with all other experiments) 
if these resonances are the lightest new (probably composite) states with a
mass comparable to the scale of the contact interactions.     

\vspace*{1cm}
\vspace*{1.5cm}
\noindent 
\rule[.1in]{17cm}{.002in}

\noindent

\begin{flushleft} CERN-TH/97-104 \\
May 1997
\end{flushleft}
\vfill\eject
%\pagestyle{empty}
%\clearpage\mbox{}\clearpage

\setcounter{page}{1}
\pagestyle{plain}

\subsection*{Introduction}

In this paper we will try to discuss the nature of the physics that could 
explain the anomalous excess observed by the two collaboration ZEUS and H1
 at HERA \cite{hera}.
An explanation could be the real  production of a light squark or a scalar 
lepto-quark  \cite{altarelli,dreiner,lq,rizzo,babu,blum,ellis} with a small 
coupling with the electron and down quark. In the squark case one
  assumes  a  $R-$parity violating scenario \cite{barb}.  
However, even if this seems the most  realistic  interpretation,  
the   possibility  of new  four 
fermion contact
 interactions \cite{altarelli,barger} cannot be ruled out, in particular if 
the existence of a peak around $200~GeV$ is not confirmed by the experiments.
These interactions are severely constrained both from atomic parity violation 
and from collider physics at the Tevatron and LEP 2 
 \cite{altarelli,barger,opal,cdf}.
As a result, only few helicity combinations are allowed and the nature of the
 physics  that could generate them is very restricted. We will see that it is
 not easy to find a simple scenario. 
First we analize the possibility that the contact terms come from new neutral 
 currents; we will see that they cannot give a satisfactory   interpretation
of the  helicity structure of the contact terms that appears to emerge when 
all the constraints are taken into account.
On the contrary, all the lepto-quarks of the adjoint of a global 
 $SU(8)\times SU(8)$ (with spin 1) contained in a $SU(16)$ global symmetry,
  could potentially  give the correct  contact terms, 
with the needed helicity structure and 
the correct overall and relative  signs.

\subsection*{Contact interactions}   
The four fermion interactions are usually parametrized in  
the following    way \cite{pesk}
\be 
 \eta_{ij} {4 \pi \over \Lambda^2}
 \bar \psi_i \gamma^\mu \psi_i ~ \bar \psi_j \gamma_\mu \psi_j 
\ee
where $i,j=e_L,e_R,u_L,u_R,...$  are  left handed and right handed components   
of the standard fermions.
 
Adding   new physics to the standard model lagrangian can potentially 
 affect the 
precision measurements at the $Z_0$ peak \cite{precis}, in particularly if this
 new physics     breaks the electroweak symmetry.
For example, a $SU(2)$ violating contact term $\bar e_R u_L ~\bar u_L e_R$
 (without a similar  $\bar e_R d_L ~\bar d_L e_R$ term) can arise  
if the fermions 
exchange a scalar $SU(2)$ doublet, and the two component (up and down) have
 very   different mass. However  if we we split the masses in this scalar 
doublet, 
the electroweak parameter $\varepsilon_1$ would get a significant contribution
 which roughly goes as 
\be
{\alpha\over 4 \pi s_W^2}{ (M_u-M_d)^2\over M_W^2  }
\ee
($M_u,M_d$ are the masses of the up and down scalar in the doublet).
Since $\varepsilon_1$ is in agreement with the Standard model within an
 accuracy of the per mill level, we easely realize that the splitting could be
 at most  of the order of $w-$boson mass which is certainly not enough to 
explain a difference in the scale of the two contact terms of the order of the 
$TeV$ (if we require that the down quark contact interaction is suppressed
 compared with the up quark one).  
In general,  a $SU(2)$ breaking operator  would be suppressed by a power 
of the Higgs $vev$  $v_{HIGGS}^2$  (if we want chirally invariant contact 
interactions   we have only an even number of fermion $SU(2)-$doublets and
 scalar doublets),
 which give 
 the scale of the contact interaction 
\be 
 v_{HIGGS}^2\over \Lambda^4.
\ee
Thus the scale $\Lambda$ has to be much closer to the weak scale and
 should be visible.

Hereafter 
we will assume that the  physics responsible for the new contact interactions 
 preserves the $SU(2)\times U(1)$ symmetry.

Previous analyses \cite{altarelli,barger}  have shown that  only few 
 four fermion 
contact interactions  could explain the excess observed at HERA: 
limits coming from the tevatron in similar processes ($q~\bar q \rightarrow
 e^+~e^-$) and atomic parity violation are strong.  
HERA  suggests  the following flavor and helicity content (see ref.
\cite{altarelli,barger} for a more detailed discussion) 
\be 
\bar e_L \gamma^\mu e_L~~ \bar u_R \gamma_\mu u_R,~~~
 \bar e_R \gamma^\mu e_R ~~\bar u_L \gamma_\mu u_L
\ee  
in order to get a sizable contribution in the relevant kinematical region,
while the tevatron suggests that the other helicity contribution
 ($LL$ and $RR$  have to be  suppressed).
To avoid  
 the constraints from atomic parity violation we preserve parity, 
 and adding 
$SU(2)$ gauge invariance,  we are obliged to introduce
 the down quark contribution
\be 
\bar e_L \gamma^\mu e_L \bar d_R \gamma_\mu d_R,~~~
 \bar e_R \gamma^\mu e_R \bar d_L \gamma_\mu d_L.
\ee  
 The full lagrangian becomes\footnote{Note that four fermion contact
 interactions that induce charged current processes are strongly constrained
\cite{alta2}.} 
\ba\label{lagrangian}
\mathrm{i}L= {4 \pi \mathrm{i}\over (3~TeV)^2}&&(
\bar L_L \gamma^\mu L_L (\bar u_R \gamma_\mu u_R+\bar d_R \gamma_\mu d_R)
\nonumber\\ 
&&+
 \bar e_R \gamma^\mu e_R~ \bar Q_L \gamma_\mu Q_L).\nonumber\\
\ea 
where $\bar L_L \gamma^\mu L_L$ is the  $SU(2)$ invariant combination of two 
lepton doublets (the same for $Q_L$).
 This appears to be the most realistic  scenario 
 that could explain the HERA  excess:
all the terms have the same scale $3~TeV$ to avoid parity violation and to 
preserve gauge invariance, and the overall  phase  in front of the parenthesis
is $+\mathrm{i}$, otherwise this term would constructively interfere 
with the standard model amplitude in $\bar q + q\rightarrow e^+ + e^-$,
 leading to  an unacceptable rate of 
production   of lepton pairs at the TEVATRON. 
  
\subsection*{The nature of the new physics}
In this section we discuss the nature of the physics that could generate 
the contact interactions of the type in eq.(\ref{lagrangian}).
If these interactions are mediated by new heavy (probably composite)
 resonances 
 then it should be possible 
to describe the interaction as  the product of the matrix elements 
of  operators $J$ as below 
\ba
\label{matrix}
& &{1\over \Lambda^2} <q|J_i |q>   <e| J^i |e> \\
 & or& \nonumber\\
\label{matrix2}
& &{1\over \Lambda^2} <e| J_i |q>  <q| J^i  |e> 
\ea
where  the sum over $i$  takes into account  any possible lorentz 
structure of the operators $J$ (scalar,...).
Firstly we analyze the former case when the interaction is due to neutral
 currents   {\it i.e.}   this new  interaction 
 is mediated by the exchange  of few neutral  particles.
A realistic  consequence of this scenario  is the existence of the 
 following similar  interactions  
\ba
\label{matrix3}
& &{1\over \Lambda^2} <e_{L,R}|J_i |e_{L,R}>   <e_{L,R}| J^i |e_{L,R}> \\
 & and & \\
& &{1\over \Lambda^2} <q_{L,R}| J_i |q_{L,R}>  <q_{L,R}| J^i  |q_{L,R}>. 
\ea
Then we get a relation 
 between the scales of  contact interactions of different processes (purely
 leptonic processes $\Lambda_{ee}$, mixed $\Lambda_{eq}$, etc.):
\be
 \Lambda^2_{eq}\simeq \Lambda_{ee} \Lambda_{qq} 
\ee

This relation has to be compared with the limits \cite{opal,cdf}
coming from  LEP2 and TEVATRON
(to obtain roughly the limit on the parity conserving 
 $\Lambda^{ll}_{LL+RR}$ we have multiplied 
 $\Lambda^2=\sqrt{2} \Lambda_{RR}^2$, see ref.\cite{opal}).
Namely\footnote{Assuming  lepton universality.}
\be 
\Lambda^{ll}>\sim 4.4~TeV ~~~ \Lambda^{qq} >\sim 1.6~TeV 
\ee 
 and applying the above relation we also get 
\be \label{lep2}
              5.6~TeV>\sim\Lambda^{ll} >\sim 4.4~TeV.
\ee
 for $\Lambda^{qq}=1.6~TeV$
and 
\be
              6.9~TeV>\sim\Lambda^{ll} >\sim4.4~TeV.
\ee
for $\Lambda^{qq}=1.3~TeV$

In practice this new contact interaction is very close to the limit of both 
 LEP 2 and the TEVATRON and will be certainly seen in the next future.

These constraints become even harder  to avoid when   
we also try   to obtain  the helicity structure 
in eq.(\ref{lagrangian}), which is rather unnatural and 
unclear  in the context of pure  neutral currents \cite{contact,godfrey}:
the existence of $LR$ and $RL$ helicity without the $LL$ and $RR$ one, 
requires at least two neutral bosons, the first coupled only to $e_L$ and $q_R$
the second only to  $e_R$ and $q_L$; these couplings, their signs 
 and the masses of these bosons have to be chosen in order to guarantee the 
compatibility with the constraints (\ref{lep2}) and at the same time 
 have to give   the correct  lagrangian (\ref{lagrangian}).
\\
Therefore we believe 
that purely  neutral currents cannot  explain the HERA excess.    
On the contrary (as will become clearer after), new   heavy lepto-quarks 
 could more naturally explain the helicity  structure of these contact terms 
in (\ref{lagrangian}) and at the same time 
 the absence of similar  signals in the $e^-~~e^+\rightarrow
e^-~~e^+$ and $q~~\bar q\rightarrow q~~\bar q$ channels.

First, let us analyze the possibility of a new scalar lepto-quark exchange.
A scalar $SU(2)-$doublet  which interacts with the  electron and the up quark
 through  a Yukawa interaction $\lambda \Phi \bar e_L u_R $ will introduce the 
 amplitude      
\be 
 -\lambda^2\bar e_L u_R { \mathrm{i}\over P^2- M_{\Phi}^2}  \bar u_R e_L
 \ee
 and  after a Fierz rearrangement (and neglecting $P^2$) can be written 
\be 
 -\lambda^2 {\mathrm{i}\over M^2_\Phi} \bar e_L \gamma^\mu e_L ~~~  
\bar u_R \gamma_\mu u_R.
 \ee
Note that the  phase of this amplitude is $- \mathrm{i}$, with  
opposite sign with respect (\ref{lagrangian}). 
This is a general result for a scalar exchange, the Yukawa coupling 
enters  the amplitude through the  square $|\lambda|^2$ and the sign 
is only determined by the overall phase of the propagator of the internal 
particle  which is $-\mathrm{i}/M^2_\Phi$ for a scalar.
This  gives a constructive interference with the standard model amplitude
at the TEVATRON (and destructive at HERA): 
it increases  the rate of production of a pair of leptons at the TEVATRON 
to  an unacceptable level (if the size is large enough to explain HERA).

We therefore consider the possibility of a heavy   spin 1 particle
exchange (much heavier than the lower bound coming from the TEVATRON
 direct searches).
In such case, we need the following set of currents
\ba\label{current}
 J^\mu&=&\bar e_L \gamma^\mu (u_R)^c, ~\overline{ (e_R)^c} \gamma^\mu u_L\\
   & & \bar e_L \gamma^\mu (d_R)^c, ~\overline{ (e_R)^c} \gamma^\mu d_L
\nonumber
\ea 
in order to satisfy all the constraints  from  gauge invariance, atomic parity 
violation, LEP 2  and TEVATRON.   
$SU(3)\times SU(2) \times U(1)$ 
gauge invariance implies  that the spin 1  resonances coupled to the 
above  currents have to be $(3,2,5/6)$ and     $(\bar 3,2,-1/6)$ (plus their 
hermitian conjugate).
Obviously these currents do not commute with the 
$SU(3)\times SU(2) \times U(1)$ generators, since they carry  color and charge
 and the commutation relations 
\be\label{commutators}
\left[T^a_{SM},T^b_{NEW}\right]= f^{abc} T^c_{NEW} ~~~~~T_{SM}\in SU(3)
\times SU(2) \times
 U(1)
\ee
( where the $f^{abc}$  
are fixed by the quantum numbers $(3,2,5/6)$ and     $(\bar 3,2,-1/6)$ of
 the  currents $J^\mu$).
 
The obvious  origin for new (conserved) currents is the 
existence of a
new symmetry involving the fermions: a global (or local) 
 transformation of the fields  is associated 
with a set of Noether currents $J^\mu$, which are conserved if this  
transformation is
 a  symmetry of the lagrangian, {\it i.e.} the action is invariant under 
this transformation. 
Let us  see if our currents can be identified as the
 Noether  currents of a global (or local) symmetry.
If so, they  satisfy a Lie algebra which can be represented 
by  some matrices which transform as the adjoint of a symmetry group
\be
 \left[T^a_{NEW},T^b_{NEW}\right]=f^{abc} T^c_{NEW}
\ee
where the $T_{NEW}$ contains the generators of our lepto-quark plus additional 
 generators needed to close the algebra.
Since all our lepto-quarks carry some charges, none of them  can
 be identified with 
the Cartan subalgebra which  is neutral, we have to add new generators to
 close the algebra, and  one can easely verify that the
minimal scenario is  embedding  the above lepto-quarks into the 
 {\bf 45} adjoint of $SO(10)$ (since we need two type of
 lepto-quark, with hypercharge $-5/6$ and $1/6$).
 
Now,   if we embed  all  the first generation of standard fermions into
 a {\bf 16}  of this group 
we easely realize that the same lepto-quarks needed to generate the contact 
terms at HERA will make the proton disastrously 
unstable.  To avoid this problem, one  could introduce the exact 
conservation  of a quantum number that forbids  proton decay
 and would be forced to   enlarge the representation of the standard
 matter to include new fermions (carrying baryon number);
they should also   have a mass very  close to weak  scale due to their chiral 
nature.
Therefore, in the next section,   we will consider a larger global 
group  to embed our lepto-quarks, without additional light fermions and still 
keeping the proton stable. 

\subsection*{$SU(8)\times SU(8)$, $SU(16)$ and $SU(15)$ global symmetries }
We will discuss the simplest  scenario 
with a $SU(8)\times SU(8)$ global symmetry, 
plus  a discrete 
symmetry $P$  which exchanges the left-handed and right-handed fermions
(all these generators  are  contained in $SU(16)$);
 one can apply similar arguments for 
$SU(15)\supset SU(7)\times SU(8)$.
  
We divide the 16 fermions\footnote{Including the right-handed neutrino.}
 in two multiplets of 8 fermions:  the first contains all the left-handed 
quarks, the right-handed electron and the right-handed neutrino; the second
contains all the right-handed quarks and  the left-handed leptonic doublet.
Each $SU(8)$ acts on  its  multiplet.

Now we will write a mass term for all the spin 1 boson of the adjoint of this 
group:  first, we introduce an invariant mass term $M_0^2$ which does not
 break  the whole symmetry group. All the bosons are degenerate with mass
 $M_0$.  Then we  break the group $SU(8)\times SU(8)$ without breaking the 
 discrete symmetry $P$: each $SU(8)$ is broken into 
$SU(6)\times SU(2)\times U(1)$;  therefore 
 we expect that all the bosons of the unbroken generators remain
 with mass $M_0$ since they would be massless for $M_0=0$.
On the contrary we expect that the 
 broken generators would have a squared mass $M_0^2+M_1^2$.
Here we assume that the sign in front of the mass term $M_1^2$ which 
split the states in the multiplet (breaking the $SU(8)$ symmetry) is negative:
 the lepto-quarks  would be the 
lightest particles with squared mass $M_0^2-M_1^2$.

Once we have written the above masses for the particles, we will have 
that the lepto-quarks which transform as the $(6,2)$ under the unbroken 
$SU(6)\times SU(2) \times U(1)$ that acts on the left-handed quarks 
(as well as the lepto-quarks acting on the right-handed quarks) 
are the lightest and degenerate states since 
also the discrete symmetry $P$ is unbroken\footnote{These symmetries are 
obviuosly broken by gauge interactions and fermion masses.}.

These lepto-quarks transform as the $(3,2,5/6)$ and $(\bar 3,2,-1/6)$ under
 $SU(3)\times SU(2) \times U(1)$. 
 For instance, the $(3,2,5/6)$  are coupled with $e_R$ and $u_L,d_L$ and give
  the following interaction
\be 
 -\lambda^2 {\mathrm{i}\over M^2} \overline{\left(e_R\right)^c} \gamma^\mu u_L
 ~~~  
\bar u_L \gamma_\mu (e_R)^c.
 \ee 
This will give one of the needed contact terms (\ref{lagrangian}) 
after a Fierz rearrangement.
In the same way, all  the other lepto-quarks will give 
all the contact interactions (\ref{lagrangian}) 
needed to explain HERA without
 affecting all the other measurements.
Here we have not addressed the problem to generate the CKM matrix in the above 
 framework, as a simplifying (and {\it ad hoc}) 
assumption we could  say, for instance,
 that the Cabibbo 
angle is mainly due to  a mixing in the up quark sector, in which case the
 lepto-quarks  would induce FCNC only in the $D_0$ decay, where the limits 
are much  less stringent.
 A complete understanding of the flavor physics certainly 
demand further work.

\subsection*{Conclusions}  
In this paper we have tried to understand the possible nature of the 
physics that 
generates the four fermion contact interaction needed to explain the HERA
 excess. Even if the explanation of a new light resonance as a squark or 
a light scalar lepto-quark  appears  to be the simplest and more realistic 
scenario, it is not
 ruled out that the anomalous excess is due to new 
four fermion contact interaction; in particular if the cluster of events  
around $200~GeV$  observed by one of the experiments
 is not confirmed. 
  Since the number of contact terms  is very restricted and 
only few helicities are allowed,  finding a simple scenario is certainly not 
an easy task.  However, we have proceeded without prejudice, and at each step 
 we have chosen the simplier (to our opinion)  direction.   
We have emphasized  that new neutral currents cannot provide a satisfactory 
explanation, because  they should affect  $e^+~e^-\rightarrow   e^+~e^-$ and 
$ q \bar q \rightarrow q \bar q$ channels , and because the helicity structure
 needed ($LR$ and $RL$ ) can only  unnaturally  arise from purely neutral 
currents.
On the contrary it appears to us that these problems could 
find a   simplier 
explanation  if the new contact interaction come from the 
exchange of  new  lepto-quark at the $TeV$ region.

We have tried to identify the $J_\mu$  currents coupled to this new particles
 as the Noether currents
  of a new global  symmetry. 
In the minimal case of a $SO(10)$ symmetry, the corresponding 
lepto-quarks make the proton
 unstable  and additional baryonic fermions (with an unnatural pattern of
 masses) are required to  conserve 
exactly the baryonic number.        
 On the contrary, a larger symmetry, as  $SU(8)\times SU(8)\subset SU(16)$
 acting on the first generation of the standard fermions contains the 
lepto-quarks of spin 1 and with quantum number     $(3,2,5/6)$ and 
    $(\bar 3,2,-1/6)$ with respect $SU(3)\times SU(2) \times U(1)$; these 
lepto-quarks   do not make the proton unstable.
If these are the  lightest new states 
 they  would  give   the four 
fermion interactions (\ref{lagrangian}),
 with the correct  helicity structure and
 the correct  overall and relative   signs.
\subsection*{Acknowledgements} 
I would like to thank  G.Altarelli, R.Barbieri and G.Giudice for very
 interesting discussions.

  \end{document}